\documentclass{aa}

\usepackage{natbib}
\usepackage[pdftex]{graphicx}%
\usepackage{epstopdf}	
\usepackage[mathcal]{eucal}
\usepackage{amssymb} 
\usepackage{amsmath}
\usepackage[usenames]{color}

\newcommand{\ds}{\displaystyle}

\usepackage[usenames]{color}	

\long\def\jumpover#1{{}}

\newcommand{\eq}[1] {Eq.~(\ref{#1})}

\newcommand{\fig}[3]{
      \begin{figure}[ht]
        \begin{center}
        \resizebox{\hsize}{!}{\includegraphics  {#1}}
        \end{center}    
        \caption{#2}
        \label{#3}
        \end{figure} }
\newcommand{\eqn} [1] {
\begin{align}#1
\end{align}}
\newcommand{\eqna} [1] {
\begin{eqnarray}#1
\end{eqnarray}}

\usepackage[applemac]{inputenc}

\def\dLrms{\left(  {{\delta L} / {L }} \right )_{\rm rms}}
\def\teff{T_{\rm eff}}
\def\pmax{ {\cal P}_{\rm max} }
\def\vmax{ {V}_{\rm max} }
\def\dLmax{\left(  {{\delta L} / {L }} \right )_{\rm max}}
\def\numax{ {\nu}_{\rm max} }
\def\pmax{ {\cal P}_{\rm max} }

\begin{document}

\title{Amplitudes of solar-like oscillations in red giant stars}
\subtitle{Evidence for non-adiabatic effects using CoRoT observations}

\author{ R. Samadi \inst{1} 
\and K. Belkacem \inst{1}
\and M.-A. Dupret \inst{2}
\and H.-G. Ludwig \inst{3,4}
\and F. Baudin \inst{5}
\and E.  Caffau \inst{3,4}
\and M.-J. Goupil \inst{1}
\and C. Barban \inst{1}
}
\institute{LESIA, CNRS UMR8109, Observatoire de Paris, Universit\'e Pierre et Marie Curie, Universit\'e Denis Diderot, Place Jules Janssen, 92195 Meudon Cedex, France
\and Institut d'Astrophysique et de G\'eophysique de l'Universit\'e de Li\`ege, All\'ee du 6 Ao\^ut 17 - B 4000 Li\`ege, Belgium 
\and Zentrum für Astronomie der Universität Heidelberg, Landessternwarte, Königstuhl 12, D-69117 Heidelberg, Germany
\and GEPI, CNRS, Observatoire de Paris, Universit\'e Denis Diderot, Place Jules Janssen, 92195 Meudon Cedex, France 
\and Institut d'Astrophysique Spatiale, CNRS, Universit\'e Paris XI,
   91405 Orsay Cedex, France
}

\date{\today}

\titlerunning{Amplitudes of solar-like oscillations in red giant stars}

\abstract
{A growing number of solar-like oscillations has been detected in red giant stars thanks to CoRoT and {{\it Kepler}} space-crafts. In the same way as for main-sequence stars, mode driving is attributed to turbulent convection in the uppermost convective layers of those stars. 
}
{
The seismic data gathered by CoRoT on red giant stars allow us to test mode driving theory in  physical conditions different from main-sequence stars. 
}
{
Using a set of 3D hydrodynamical models representative of the upper layers of  sub- and red giant stars, we computed the acoustic mode energy supply rate  ($\pmax$). Assuming adiabatic pulsations and using global stellar models that assume that the surface stratification comes from the 3D hydrodynamical models, we computed the mode amplitude in terms of surface velocity. This was converted into intensity fluctuations using either a simplified adiabatic scaling relation or a non-adiabatic one.
}
{
From $L$ and $M$ (the luminosity and mass), the energy supply rate $\pmax$ is found to scale as $(L/M)^{2.6}$ for both main-sequence and red giant stars, extending previous results. 
  The theoretical amplitudes in velocity  under-estimate the Doppler velocity measurements obtained so far from the ground for red giant stars by about 30\,\%.  In terms of intensity, the theoretical scaling law based on the adiabatic intensity-velocity scaling relation results in an under-estimation  by a factor of about 2.5 with respect to the CoRoT seismic measurements. On the other hand, using the non-adiabatic intensity-velocity relation  significantly reduces the discrepancy with the CoRoT data.   The theoretical amplitudes remain  40\,\% below, however, the CoRoT measurements.
}
{
 Our results show that scaling relations of mode amplitudes cannot be simply extended from main-sequence to red giant stars in terms of intensity on the basis of adiabatic relations because non-adiabatic effects for red giant stars are important and cannot be neglected. We discuss possible reasons for  the remaining differences.
 }


\maketitle

\section{Introduction}

Before CoRoT (launched in December 2006), solar-like oscillations had been detected for a dozen of bright red giant stars either from the ground or from  space with MOST \citep[e.g., ][]{Barban07}. 
Thanks to CoRoT and {\it Kepler}, it is now possible to detect and measure
solar-like oscillations in many more (several thousands) 
red giant stars  \citep[e.g.,][]{deRidder09,Huber10,Bedding10,Kallinger10,Stello11,Mosser12}.   
With such a large set of stars, it is possible to perform ensemble
asteroseismology by deriving scaling
relations  that relate seismic parameters to a few fundamental stellar
parameters (e.g. masses, radii, luminosities etc). These approaches
are now commonly applied to global seismic parameters, such as the
cutoff-frequency or peak frequency  \citep[e.g.,][]{Miglio09,Stello09,Kallinger10,Mosser10}. However, scaling relation is used only infrequently for mode amplitudes. The main reason for this is our poor theoretical understanding of the  underlying physical mechanisms for mode driving and damping. 

Using  a large set of red giant stars observed by CoRoT, \citet{Baudin11} have derived  scaling relations  in terms of mode lifetimes and amplitudes. These authors have found that  the scaling relation proposed by \citet{Samadi07a} for the mode amplitude significantly departs from the measured one. 
This result was recently confirmed by \citet{Huber11}, \citet{Stello11} and \citet{Mosser12}  with {\it Kepler} observations, and is easily understood by noting that \citet{Samadi07a} established the scaling for \emph{for main-sequence stars} only, and only for \emph{mode surface velocity}. Indeed, those results point out that a dedicated theoretical investigation of mode amplitudes in intensity for red giants is needed to provide an adequate theoretical background. 

Towards the end of their lives, low-mass stars greatly expand their envelope to become red giant stars. As a consequence, the low density of the envelope favours a vigorous convection such that 
excitation of solar-like oscillations occurs in a medium with very different physical conditions than encountered in the Sun. This introduces new problems about the physical mechanism related to mode driving. For instance, the higher the turbulent Mach number, the more questionable the assumptions involved in the theory  \citep{GK77,GMK94,Samadi00I,Chaplin05,Kevin10}. 

In addition, red giant stars are characterised by high luminosities and hence have relatively short convective thermal time-scales at the upper most part of their convective envelope. One can therefore expect a stronger departure from adiabatic oscillations because the perturbation of entropy fluctuations related to the oscillations dimensionally depends on the ratio $L/M$ (where $L$ is the luminosity and $M$ the mass). 
Thus, extreme physical conditions in the uppermost convective regions of red giants raise new questions about the energetic aspects of damped stochastically excited oscillations (more precisely mode driving and damping). 
In the present paper, we focus on modelling mode driving. We   derive scaling relations for red giant stars in terms of mode amplitude (in velocity and intensity) and  compare them with the available CoRoT observations. 

This paper is organised as follows: 
from a grid of 3D hydrodynamical models representative for the upper layers of red giant stars, we derive in Sect.~\ref{scaling_laws} theoretical scaling laws for the mode amplitudes in velocity (Sect.~\ref{velocity}) and in intensity (Sect.~\ref{intensity}). These  scaling laws are then compared in Sect.~\ref{comparison} with  seismic data. Finally, Sect.~\ref{conclusion} is dedicated to conclusions. 


\section{Theoretical scaling relations for mode amplitudes}
\label{scaling_laws}

In this section our objective is to compute theoretical scaling relations of mode amplitudes both in terms of surface velocity and intensity. To this end, the mode amplitude will be computed with the help of hydrodynamical 3D numerical simulations. 

\subsection{Surface velocity mode  amplitude, $v$}
\label{velocity}

The mean-squared surface velocity for each \emph{radial}  mode
is given by \citep[e.g.][and references therein]{Samadi10}
\begin{equation}
\label{v}
v^2 (\nu,r) = {\tau(\nu) \over 2} \, \frac{ {\cal P}(\nu) }{ \mathcal{M}(\nu,r) } \; ,
\end{equation}
where $\nu$ is the mode frequency, ${\cal P}$  the mode excitation rate, $\tau$ the mode life-time (which is equal to the inverse of the mode damping rate $\eta$),  $\mathcal{M}$ the mode mass, and $r$ the radius in the atmosphere where the mode velocity is evaluated.
The mode mass $\cal M$ is defined for radial modes as
\begin{equation}
\mathcal{M}(\nu,r) = { 1 \over { \vert \xi_r(\nu,r) \vert^2 }} \,  \int_{0}^{M}   \vert \xi_r(\nu,m) \vert^2 \, {\rm d} m \;, 
\label{MM}
\end{equation}
where $\xi_r$ is the radial component of the mode
eigendisplacement. 
The quantities $v$, $\cal M$ and $\xi_r$ are evaluated  at two relevant layers:
\begin{itemize}
\item the photosphere, \emph{i.e.} at  $r=R_{*}$ where $R_*$ is the stellar radius; 
\item at a layer where spectrographs dedicated to stellar seismology are the most sensitive. According to \citet{Samadi08},  for the  Sun and solar-type stars, this layer is close to the  depth where the potassium (K) spectral line is formed, that is  at the optical depth $\tau_{\rm~500~nm} \simeq 0.013$. For stars with different spectral type this layer may vary, but by an as yet unknown manner \citep[see the discussion in][]{Samadi08}. By default we therefore adopt this reference optical depth to be representative for the Doppler velocity measurements for red giant stars. This assumption is discussed in Sect.~\ref{comp_velocity}.
\end{itemize}

In \eq{v}, ${\cal P}$ and ${\cal M}$ are computed in the manner of \citet{Samadi08} using a set of 3D hydrodynamical models of the upper layers of sub- and red giant stars. However, this calculation differs from  \citet{Samadi08} in two aspects. First, instead of adopting a pure Lorentzian function for the eddy-time correlation in the Fourier domain,  we introduce, following \citet{Kevin10}, a cut-off frequency derived from the sweeping assumption. 
Second, the 3D models at our disposal have a limited vertical extent that results in an under-estimation by up to $\sim 10$\,\% of the maximum of ${\cal P}$. To take into account the driving that occurs at deeper layers we extend the calculation to deeper layers using standard  1D stellar models (see below).

 The 3D hydrodynamical models  were built with the CO$^5$BOLD code \citep{Freytag02,Wedemeyer04,Freytag12}.
All 3D models have a solar metal abundance. The chemical mixture is based on \citet{Asplund05}. 
The characteristics of these 3D models are given in Table \ref{tab:3Dmodels}. All models have a helium abundance of $Y=0.249$ and a metal abundance of $Z=0.0135$.
The 3D models S1, S2, S3, and S7 correspond to red giant stars while S4, S5 and S6 to sub-giants stars. 

For each 3D model, an associated complete 1D model (interior+surface)
is computed in such a way that the outer layers are obtained from  the
3D model \citep[see][ for details]{Samadi08} while the interior layers are  computed using the
CESAM2K code \citep{Morel08}. In these 1D models, convection is treated according to the \citet{Canuto96} local formulation of convection.
This formulation requires a prescription for the size $\Lambda$ of the strongest eddies. We assume that $\Lambda = \alpha \, H_p$ where $H_p$ is the pressure scale height and $\alpha$ a parameter adjusted such that the interior model matches the associated 3D model as detailed in \citet{Samadi08}. 
 The complete models (interior+surface) are from now on referred to as \emph{patched} models.

The characteristics of the patched models are given in Table~\ref{tab:1Dmodels}. 
 We then computed the global acoustic modes associated with each of the patched models using  the adiabatic pulsation code ADIPLS \citep{ADIPLS08}.
Finally, the mode lifetimes $\tau$ are evaluated using the measurements performed by \citet[][see Sect.~\ref{corot_data}]{Baudin11}. 

\begin{table}[ht]
\begin{center}
\begin{tabular}{ccc}
 Label & $\log g$ &   $\teff$\\
& & [K]  \\
\hline
S1  & 2.50 &  $4964~\pm~22$\\
S2   & 2.50 &  $4475~\pm~10$\\
S3   & 2.00 &  $4551~\pm~16$ \\
S4   & 3.50 &  $4931~\pm~20$ \\
S5   & 3.50 &  $5431~\pm~23$ \\
S6   & 3.50 &  $5885~\pm~16$ \\
S7  & 3.00 &  $5039~\pm~11$ 
\end{tabular}
\end{center}
\caption{Characteristics of the 3D models. $\teff$ is the effective temperature, and $g$ the surface gravity.}
\label{tab:3Dmodels}
\end{table}

\begin{table}[ht]
\begin{center}
\begin{tabular}{cccccccc}
 Label & $M$ & $\alpha$ &  $\log g$ &   $\teff$ &  $L$  & $\Delta \nu$ & $  \nu_c$ \\
 & [$M_\odot$] &  &  & [K] & [$L_\odot$] & [$\mu$Hz]  & [$\mu$Hz]\\
\hline
M1 & $3.74$ & $0.565$ & $2.51$ &  $4962$ & $172.5$ & 3.43 &63 \\
M2 & $0.98$ & $0.621$ & $2.50$ &  $4463$ & $30.4$  & 4.77 & 67\\
M3 & $4.20$ & $0.610$ & $1.99$ &  $4551$ & $444$   & 1.40 & 21\\
M4 & $1.39$ & $0.636$ & $3.53$ &  $4927$ & $5.86$  & 25.11& 637\\
M5 & $1.74$ & $0.596$ & $3.50$ &  $5392$ & $11.5$  & 23.30 &607\\
M6 & $1.73$ & $0.576$ & $3.51$ &  $5856$ & $15.9 $ & 23.30 &583\\
M7 & $2.49$ & $0.615$ & $3.00$ &  $5040$ & $39.3 $ & 9.00& 199
\end{tabular}
\end{center}
\caption{Characteristics of the 1D ``patched'' models.  $L$ is the luminosity, $M$ the mass, $\Delta \nu$ the large separation, and $\nu_c$ the acoustic cutoff-frequency.}
\label{tab:1Dmodels}
\end{table}

 Our objective is to establish a scaling for the maximum of $v$ (Eq. ~\ref{v}, $\vmax$ hereafter) as a function of stellar parameters and assuming that the mode lifetime $\tau$ is known.
 As shown by \citet{Kevin11}, the mode lifetime $\tau$ is expected to reach a plateau at a characteristic frequency, $\nu_{\rm max}$.  As we will see in Sect.~\ref{sec:pmax},  the maximum of $\left ({\cal P}/{\cal M} \right )$ also peaks at $\nu_{\rm max}$.  
Accordingly, to derive a scaling law for $\vmax$,  one needs to determine how the ratio $\left ({\cal P} / {\cal M} \right )_{\rm max}$ scales with  stellar parameters (see Sect.~\ref{sec:pmax}).

Among these parameters, apart from the classical fundamental parameters (luminosity $L$, mass $M$, effective temperature $\teff$, gravity $g$, etc), we  in addition considered the acoustic cut-off frequency $\nu_{\rm c}$ and the large frequency separation $\Delta \nu$ \citep[see e.g.][]{JCD82}, since the former is related to the properties of the surface and the latter to the mean density of the star.
These parameters scale as
\eqna{
\nu_c & =& \nu_{c,\odot} \, { {g \over g_\odot} \, \sqrt{ T_{{\rm eff},\odot} \over \teff}} \label{nuc_scaling} \\
\Delta \nu &= & \Delta \nu_{\odot} \, \sqrt{ {M  \over M_\odot } \, \left ({R_\odot \over R} \right ) ^3} \;, 
\label{deltanu_scaling}
}
where  quantities labelled with the symbol $\odot$ refer to solar values, $\nu_{c,\odot} = 5 100\,\mu$Hz \citep[see][and references therein]{Jimenez06}, and $\Delta \nu_{\odot} = 134.9\,\mu$Hz \citep{Toutain92}.
The values of $\nu_c$ and $\Delta \nu$ associated with each model are given in Table\,\ref{tab:1Dmodels}.

Finally, we stress that the characteristic frequency $\nu_{\rm max}$, at which $\tau$ reaches a plateau and ${\cal P}/ {\cal  M}$ is maximum, is related to a resonance in the uppermost layers of solar-like stars between the thermal time-scale and the modal period  \citep[see][and reference therein]{Kevin11}. This is why it scales as the acoustic cut-off frequency $\nu_{\rm c}$ in very good approximation:
\eqn{
\nu_{\rm max } = \nu_{{\rm max},\odot} \, { \nu_c \over \nu_{c,\odot} } \;,
\label{numax_scaling}
}
where   $\nu_{{\rm max},\odot} = 3 101\,\mu$Hz.

\subsection{Scaling relation for $\left ({\cal P}/{\cal M} \right )_{\rm max}$  }
\label{sec:pmax}

\begin{figure}[t]
\begin{center}
\resizebox{\hsize}{!}{\includegraphics{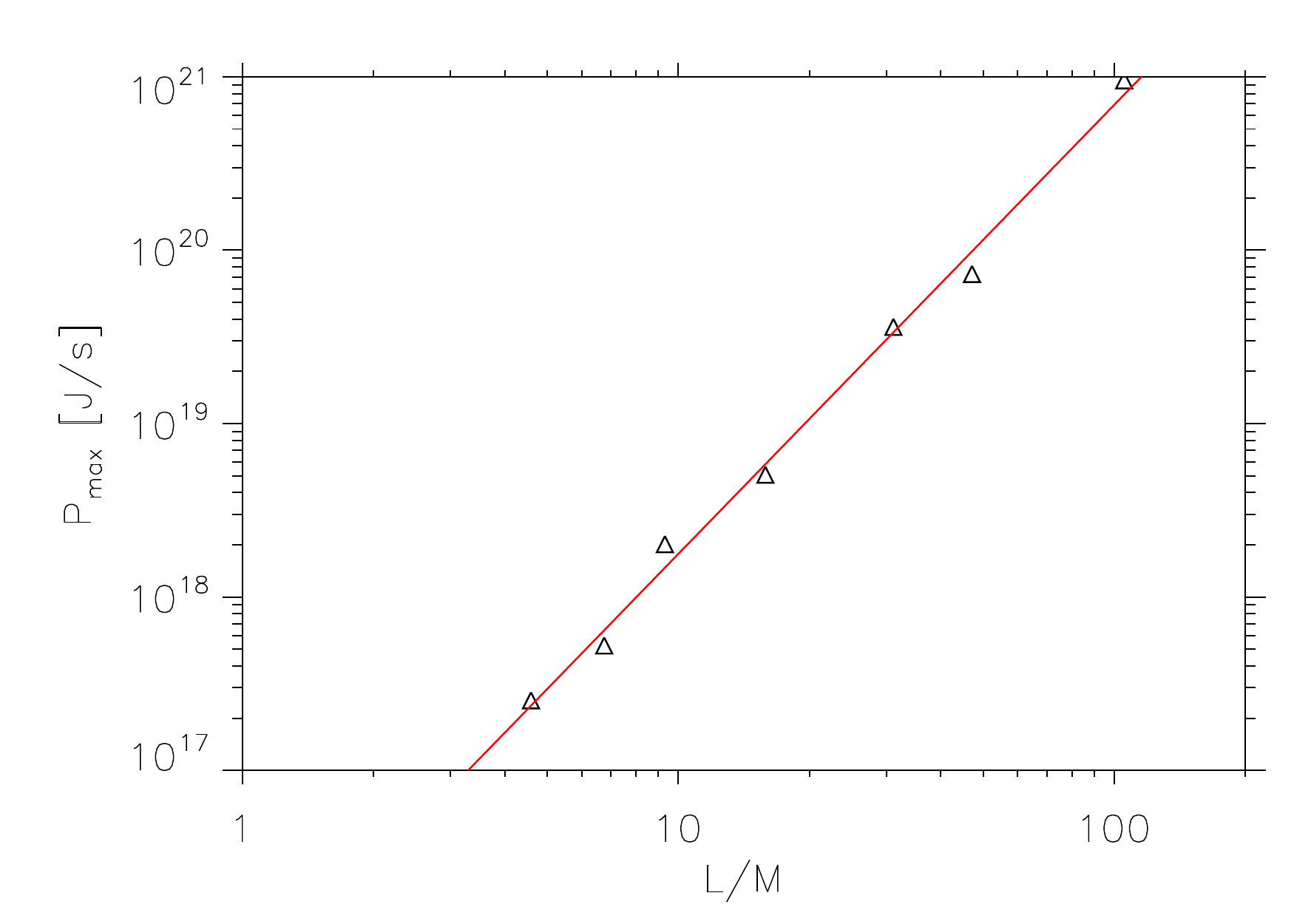}}\\
 \resizebox{\hsize}{!}{\includegraphics{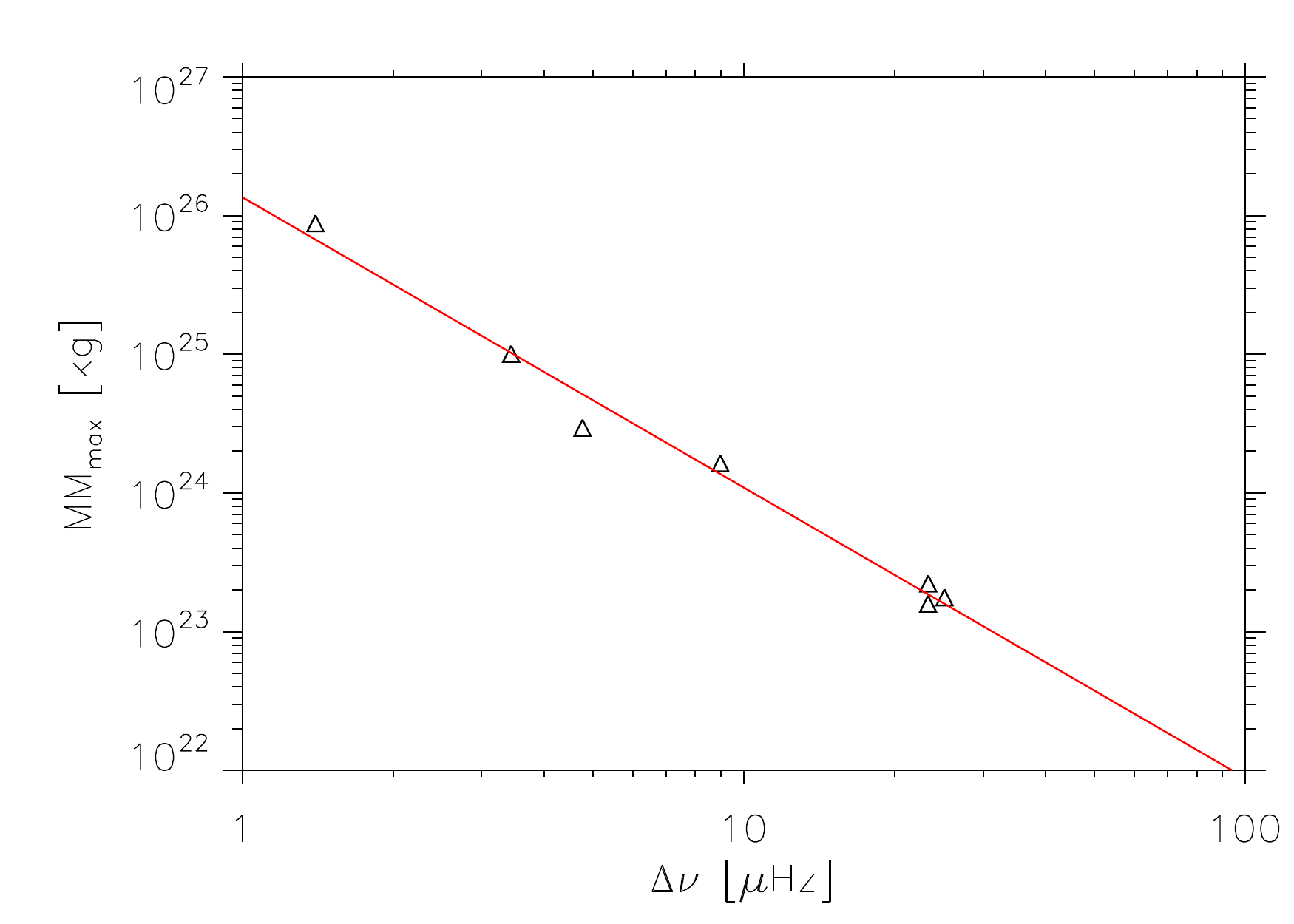}}
\end{center}    
\caption{{\bf Top:} ${\cal P}_{\rm max}$ as a function of $L/M$.  
The triangles are associated with the 3D models. The red line is a power law of the form  $ (L/M)^s$ with $s=2.6$. {\bf Bottom:} Mode mass at $\nu = \nu_{\rm max}$ (${\cal M}_{\rm max}$) as a function of the large separation $\Delta \nu$. The mode masses are  evaluated here at the layer corresponding to the optical depth $\tau_{\rm~500~nm} = 0.013$  (see text). The triangles are associated with the 3D models. The red line is a power law of the form  $(\Delta \nu/{\Delta \nu}_{\odot})^{-p}$ with $p=2.1$.}
\label{pmax}
\end{figure}

The maximum of ${\cal P}$ is plotted in Fig.~\ref{pmax} (top) as a function of the ratio $L/M \propto \teff^4/ g$.
This dependence with $\teff$ and $g$ was already highlighted and explained by \citet{Stein04} and \citet{Samadi07a} \citep[see also the review by][]{Samadi10}, and is nicely confirmed by  Fig.~\ref{pmax} (top). Indeed, 
 ${\cal P}_{\rm max}$ follows a power law of the form
 \begin{align}
 \label{eq:pmax}
{\cal P}_{\rm max} = {\cal P}_{\rm max}^0 \,  \left({L \over L_\odot} \, { M_\odot \over M}\right)^s \quad {\rm with} \quad s=2.60 \pm 0.08 \;,
\end{align} 
 where $\pmax^0 = \left ( 4.2 ^{+1.0}_{-0.8} \right )\, \times 10^{15} $J/s. The maximum of ${\cal P}$ is found to peak at a frequency close to $\nu_{\rm max}$. We note also that the value of the exponent  and the constant $\pmax^0$  in \eq{eq:pmax} are compatible with the results of \citet{Samadi07a} established on the basis of a small set of 3D models of the surface layers of main-sequence (MS) stars. We thus confirm the validity of this relation from MS to red giant stars.

We turn now to the  mode mass, ${\cal M}$. Because we aim to compare theoretical mode velocities with measurements made from the ground with spectrographs dedicated to stellar seismology, we  evaluate ${\cal M}$ at  the optical depth  $\tau_{\rm~500~nm} = 0.013$ (see Sect.~\ref{velocity} and \citet{Samadi08}).
For a given model,  the mode mass (${\cal M}$)  decreases rapidly with $\nu$, but above a characteristic frequency close to $\nu_{\rm max}$ it decreases more slowly.
Although  ${\cal M}$ does not have a minimum, we found that, as ${\cal P} $, the ratio $\left ({\cal P} / {\cal M} \right )$ reaches a maximum close to  $\nu_{\rm max}$, which scales as given by \eq{nuc_scaling} and \eq{numax_scaling}. Therefore, we evaluate ${\cal M}$ at $\nu = \nu_{\rm max}$. From now on we label this quantity as ${\cal M}_{\rm max}$.

Among the different stellar parameters mentioned in Sect.~\ref{velocity}, a clear correlation of ${\cal M}_{\rm max}$ is found with $g$, $(L/M)$, $\nu_c$ or $\Delta \nu$. However, the more pronounced correlation is found with $\Delta \nu$. We therefore adopt the scaling with $\Delta \nu$. The variation of  ${\cal M}_{\rm max}$ with $\Delta \nu$ is shown in Fig.~\ref{pmax} (bottom). 
${\cal M}_{\rm max}$ can be nicely fitted by a power law of the form
\begin{align}
 \label{M_min}
{\cal M}_{\rm max} = {\cal M}_{\rm max}^0 \, \left({\Delta \nu \over {\Delta \nu } _{\odot}}\right)^{-p} \quad {\rm with} \quad p = 2.1 \pm 0.1 \;,
\end{align}
where ${\cal M}_{\rm max}^0= \left (4.5 ^{+1.8}_{-1.3} \right )\, \times \,10^{21}$~kg, and $\Delta \nu$ is given by the scaling relation of \eq{deltanu_scaling}.

By using \eq{eq:pmax} and \eq{M_min}, the maximum of the ratio ${\cal P}/{\cal M}$ then varies according to: 
\begin{align}
\label{scaling_PM}
 \left( {\cal P}/{\cal M} \right)_{\rm max} = (\pmax^0/{\cal M}_{\rm max}^0)   \,\left({L \over L_\odot} { M_\odot \over M} \right)^s \, \left ( { {\Delta \nu} \over  {\Delta \nu}_{\odot} } \right )^{p} \;.
\end{align}

\subsection{Scaling relation for $\vmax$}
\label{sec:vmax}

Equation (\ref{scaling_PM}) now permits us to proceed by considering the scaling law for mode amplitudes, in terms of surface velocities. The maximum of the mode surface velocity, by using \eq{scaling_PM} together with \eq{v}, reads 
\begin{equation}
\vmax =  v_0 \, \sqrt{  {{\tau_{\rm max} }  \over { \tau_{0}}} \,  \left ( {L \over L_\odot} { M_\odot \over M}   \right )^{s} \, \left ({ {\Delta \nu} \over {\Delta \nu}_{\odot}} \right )^{p} } \; .
\label{eq:vmax}
\end{equation} 
where $\tau_{\rm max}$ is the characteristic  lifetime at $\nu = \numax$, and 
\begin{equation}
v_0= \sqrt{\ds {\tau_0 \over 2} \, \left ({\pmax^0} \over {{\cal M}_{\rm max}^0} \right ) } \, , 
\end{equation} 
with $\tau_0$ a reference mode lifetime whose values are arbitrary fixed to the lifetime of the solar radial modes at the peak frequency, that is  $\tau_0= 3.88$~days. Accordingly, we have $v_0= 0.41~$m/s.

It is worthwhile to note that our scaling relation (Eq.~\ref{eq:vmax}) differs from the result of \cite{Kjeldsen11}. This is  explained by the fact that the postulated relation  of \cite{Kjeldsen11} for mode amplitudes in velocity (their equation 16) does not take  the mode masses into account, while this is definitively necessary as seen in \eq{v}. 

\subsection{Scaling relation for bolometric amplitude}
\label{intensity}

The instantaneous bolometric  mode amplitude  is  deduced at the photosphere according to \citep{Dziembowski77a,Pesnell90}   
\begin{equation}
\frac{\delta L (t)}{L}  = 4\, \frac{\delta \teff(t)}{\teff} 
+ 2\, \frac{\delta R_* (t)}{R_*} \;,
\label{dL}
\end{equation}
where $\delta L(t)$ is the mode Lagrangian (bolometric) luminosity perturbation, $\delta \teff(t)$ the  effective temperature fluctuation, and $\delta R_*(t)$ the variation of the stellar radius.

Since the second term of \eq{dL} is found negligible in front of $\delta \teff(t)$, one obtains the rms bolometric  amplitudes according to
\begin{equation} 
\left(  {{\delta L} \over {L }} \right )_{\rm rms}    =   4  \left (\ds \frac{\delta \teff}{\teff} \right )_{\rm
  rms} \label{dLrms0} \;,
\end{equation}
where the subscript rms denotes the root mean-square. 

We now need  a relation between $\left(  {{\delta \teff }  / \teff } \right )_{\rm rms}$ (or equivalently $\dLrms$) and the rms mode velocity $\vmax$.  
For convenience we introduce the dimensionless coefficient $\zeta$ defined according to 
\eqn{
\dLrms =  4  \left (\ds \frac{\delta \teff}{\teff} \right )_{\rm
  rms} =  \zeta \,  \dLrms^\odot \, \left ( { v_{\rm rms} \over  v_\odot }   \right ) \;,
\label{dLrms}
}
where  $\dLrms^\odot = 2.53$~$\pm$0.11~ppm is the maximum of the solar bolometric mode amplitude  \citep{Michel09}, $\teff^\odot=5777$~K the effective temperature of the Sun, and  $v^\odot_{\rm rms}=18.5~\pm~1.5$~cm/s the  maximum of the solar mode (intrinsic)
surface velocity evaluated at the photosphere as explained in
\citet{Samadi09b}.

The quantity $\zeta$ in \eq{dLrms} is defined at an arbitrary layer, which is generally the photosphere (i.e. at $r=R_{*}$). Accordingly, we must evaluate the velocity and hence the mode mass $\cal M$ at that layer. This implies  the following scaling for $\cal M_{\rm max}$:
\begin{align}
 \label{M_min_photo}
{\cal M}_{\rm max,*} = {\cal M}_{\rm max,*}^0 \, \left({\Delta \nu \over \Delta \nu^{\odot}}\right)^{-p_*} \;,
\end{align}
where $ p_* = 2.0 \pm 0.10$ , ${\cal M}_{\rm max,*}^0= \left ( 8.0^{+2.8}_ {-2.1}  \right ) \times 10^{21}$~kg and   $\Delta \nu$ is given by the scaling relation of \eq{deltanu_scaling}.

Combining \eq{dLrms} with  \eq{eq:vmax} gives the scaling for the bolometric amplitude
\eqna{
 \left(\frac{\delta L}{L}\right)_{\rm max}&  = & \zeta \, \left(\frac{\delta L}{L}\right)^\odot_{\rm rms} \nonumber \left ( {v_{0,*} \over v^\odot_{\rm rms} } \right ) \,\\ & & \times  \sqrt{  {{\tau_{\rm max} }  \over { \tau_{0}}} \,  \left ( {L \over L_\odot} { M_\odot \over M}  \right )^{s} \, \left ({ \Delta \nu \over \Delta \nu^{\odot}} \right )^{p_*} } \;,
\label{dLmax}
}
where $v_{0,*} \equiv \sqrt{\ds {\tau_0 \over 2} \, \left ({\pmax^0} \over {{\cal M}_{\rm max,*}^0} \right ) } = 0.31 $~m/s.

\subsubsection{Adiabatic case}

Within the adiabatic approximation, it is possible to relate the mode surface velocity to intensity perturbations \cite[e.g.,][]{Kjeldsen95}; this give:
\eqn{
\zeta_{\rm K95} = \sqrt{ \teff^\odot  \over   \teff }  \;,
\label{zeta_K95}
}
which  assumes that the modes are quasi-adiabatic, but not only. It supposes that  the modes propagate at the surface where they are measured. This approximation is not valid in the region where the modes are measured since in this region they are evanescent.  Furthermore, it assumes an isothermal atmosphere. 
A more sophisticated quasi-adiabatic approach has been proposed by \citet{Severino08}. The authors went beyond the approximation of isothermal atmosphere by taking into account the temperature gradient as well as the fact that the intensity is measured at constant instantaneous  optical depth. 
Both effects are  taken into account by the method described in Sect.~\ref{nad}, which in addition considers non-adiabatic modes.

We present in  Fig.~\ref{intensity-velocity} $\zeta_{\rm K95}$ as a function of $(L/M)$. 
The adiabatic coefficient remains almost constant for the type of stars investigated here (sub- and 
red giant stars). This is obviously because $\zeta_{\rm K95}$ varies as the inverse of the square root of $\teff$. 

\subsubsection{Non-adiabatic case}
\label{nad}

We also computed  $\zeta$  using the MAD non-adiabatic pulsation code \citep{Grigahcene05}. 
This code includes the time-dependent convection (TDC) treatment described in \cite{Grigahcene05}. 

This TDC formulation involves a free parameter $\beta$, which takes complex values  and enters the
perturbed  energy equation. This parameter was introduced to prevent  the occurrence of  non-physical spatial oscillations in the eigenfunctions \cite[see][ for details]{Grigahcene05}. To constrain this parameter we used the scaling relation between the frequency of the maximum height in the power spectrum ($\nu_{\max}$) and the cut-off frequency ($\nu_{\rm c}$). When scaled to the Sun, one can use this scaling to infer $\nu_{\rm max}$ for the models we used and the parameter $\beta$ is then adjusted so that the plateau (or depression) of the computed damping rates coincides \citep[see][]{Kevin12}. 

Note also that TDC  is a non-local  formulation of  convection and is based on the \cite{Gabriel96} formalism as explained in \citet{MAD06b} and \citet{MAD06c}. In this framework, non-local parameters related to the convective flux ($a$) and the turbulent pressure ($b$) are chosen in the same way as in \citet[][see their Eqs. (17) and (18), see also \citet{MAD06a}]{MAD06b} so that it fits the solar 3D numerical simulation. This calibration results in $a=10.4$ and $b=2.9$ (assuming a mixing-length parameter $\alpha=1.62$)

\begin{figure}[t]
\begin{center}
\resizebox{\hsize}{!}{\includegraphics{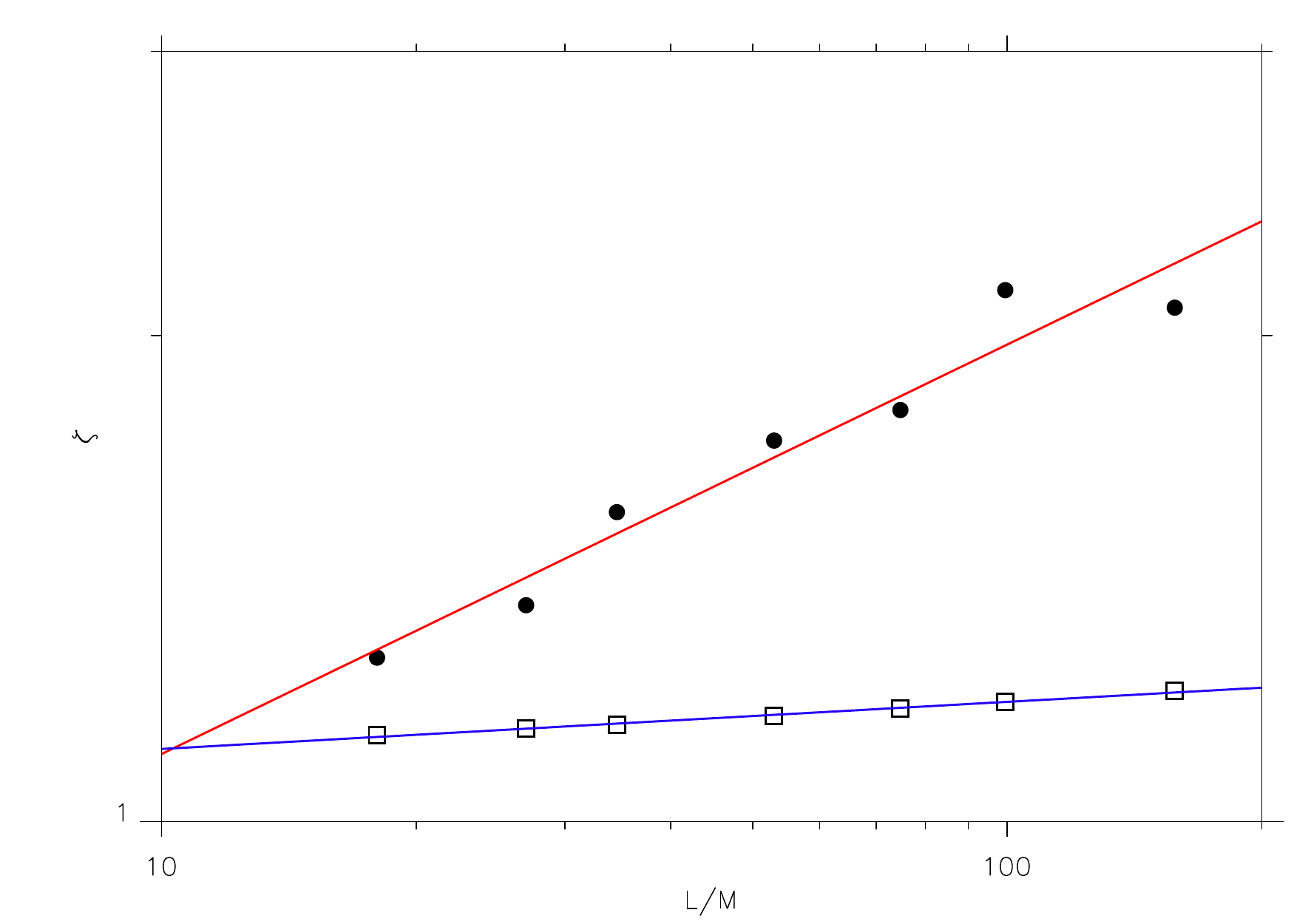}}\\
\end{center}    
\caption{Coefficient $\zeta$ (see Eq.~\ref{dLrms}) as a function of $L/M$ for sub- and red giants. The filled circles correspond to the values, $\zeta_{\rm nad}$, obtained with the MAD non-adiabatic pulsation code (see details in the text). The empty squares correspond to the adiabatic coefficient  \citet{Kjeldsen95} (see Eq.~\ref{zeta_K95}). The red line corresponds to a power law of the form $\zeta_0 \, (L/M)^k$ with $k$=0.25. Both intensity-velocity relations $\zeta$ have  been calibrated so that for the Sun   $\zeta=1$ (see text).}
\label{intensity-velocity}
\end{figure}

For sub- and red giant stars ($L/M \gtrsim 10 \, L_\odot/M_\odot$), the non-adiabatic intensity-velocity relation  obtained with the MAD code can quite well be fitted by a power law of the form
\eqn{
\zeta_{\rm nad} = \zeta_0 \, \left ( {L \over L_\odot} \, { M_\odot \over M} \right )^k \;,
\label{zeta_nad}
}
where $k = 0.25 \pm 0.05$ and $\zeta_0=0.59 \pm 0.07$. 
 For main-sequence stars ($L/M \lesssim 10 \, L_\odot/M_\odot$), $\zeta_{\rm nad}$ remains almost constant (not shown). For the Sun, we find $\zeta_{\rm nad} \simeq 0.95$, which is close to the value  expected by definition for the Sun. Therefore, we are then led to multiply $\zeta_{\rm nad}$ by only  a factor 1.05 so that, for the Sun, theoretical $\dLmax$ matches the helioseismic measurements. The result is shown in Fig.~\ref{intensity-velocity} for the  sub- and red giant stars ($L/M \gtrsim 10\, (L_\odot/M_\odot)$). The non-adiabatic coefficient increases  rapidly with increasing $(L/M)$ while  $\zeta_{\rm K95} $ remains almost constant.
Hence, the higher $(L/M)$, the larger the difference between the non-adiabatic and the adiabatic coefficient ($\zeta_{K95}$).

\section{Comparison with the observations}
\label{comparison}

We compare in this section theoretical mode amplitudes with seismic measurements made from the ground in terms of Doppler velocity (Sect.~\ref{comp_velocity}) and from space by CoRoT in terms of intensity (Sect.~\ref{comp_intensity}). We recall that computing the theoretical mode amplitudes requires   knowledge of  $\tau_{\rm max}$  (see Eqs.~\ref{eq:vmax} and \ref{dLmax}), which is obtained from a set of CoRoT targets as explained in Sect. \ref{corot_data}. 

\subsection{The CoRoT data set}
\label{corot_data}

\citet{Baudin11} have measured the mode amplitudes for $360$ CoRoT red giant targets.  Among those targets, many  show very narrow 
peaks, close to the frequency resolution of the spectrum, while the others have resolved peaks. About  65\,\%  of those targets have a highest mode whose width is sufficiently broad to be fitted with a Lorentzian profile. For those targets, the height of the highest mode, $H_{\rm max}$, and its lifetime  $\tau_{\rm max}$ are thus derived from the fit procedure. However, it is not excluded that some modes with a width more narrow than the frequency resolution  may have  been fitted with a Lorentzian profile because of the low signal-to-noise ratio. 
To exclude those modes, we only considered modes with a width $\Gamma_{\rm max} = 1/(\pi \tau_{\rm max})$ broader than twice the frequency resolution of the spectra (which is $0.081~\mu$Hz). 
This subset represents about 170 targets for which we have an estimate of the mode lifetime ($\tau_{\rm max}$)  at the peak frequency.
For each target of this subset, the maximum of the mode amplitude in intensity  ($A_{\rm max}$) was obtained according to the relation  $A_{\rm max}= {\ds \sqrt{  H_{\rm max} / \tau_{\rm max}}}$. Finally, a bolometric correction was applied in the manner of \cite{Michel09} to convert the apparent intensity fluctuation  $A_{\rm max}$ into a bolometric amplitude $(\delta L/L)_{\rm max}$.

\subsection{Maximum velocity amplitude ($\vmax$)}
\label{comp_velocity}

The mode amplitude in terms of velocity is given by \eq{eq:vmax}. Calculating   $\vmax$ requires to know the mode  life time $\tau_{\rm max}$ at the peak frequency. 
We used the values of  $\tau_{\rm max}$ available  for our set of CoRoT targets (see Sect.~\ref{corot_data}).  We also determined the ratio ${L / M}$ as well as $\Delta \nu$.
The luminosity and mass of these targets are unknown. However, \citet{Baudin11} have proposed to derive an estimate of the ratio $L/M$ using the following scaling:
\eqn{
{L \over M}  \propto { {\teff^{7/2}} \over {\nu_{\rm max}} } \;,
\label{L_M}
}
where $\nu_{\rm max}$ is the frequency of the maximum mode height
$H_{\rm max}$ and  $\teff$ is determined from photometric
broad-band measurements as explained in \citet{Baudin11}. Note that
the scaling law of \eq{L_M} assumes that  $\nu_{\rm max}$ scales as
$\nu_c$, which scales as $g/\sqrt{\teff}$ (see
Eq.~\ref{nuc_scaling}). Concerning $\Delta \nu$, 
 as first established by  \citet{Stello09}, \citet{Hekker09} and
  \citet{Kallinger10}, there is a clear scaling relation between this quantity and $\numax$. We derived  
 this quantity here according to the relation derived by \citet{Mosser10} from a large set of CoRoT red giant stars:
\eqn{
\Delta \nu = 0.280 \, \nu_{\rm max}^{0.747} \; .
}

\begin{figure}
    \begin{center}
        \resizebox{\hsize}{!}{\includegraphics  {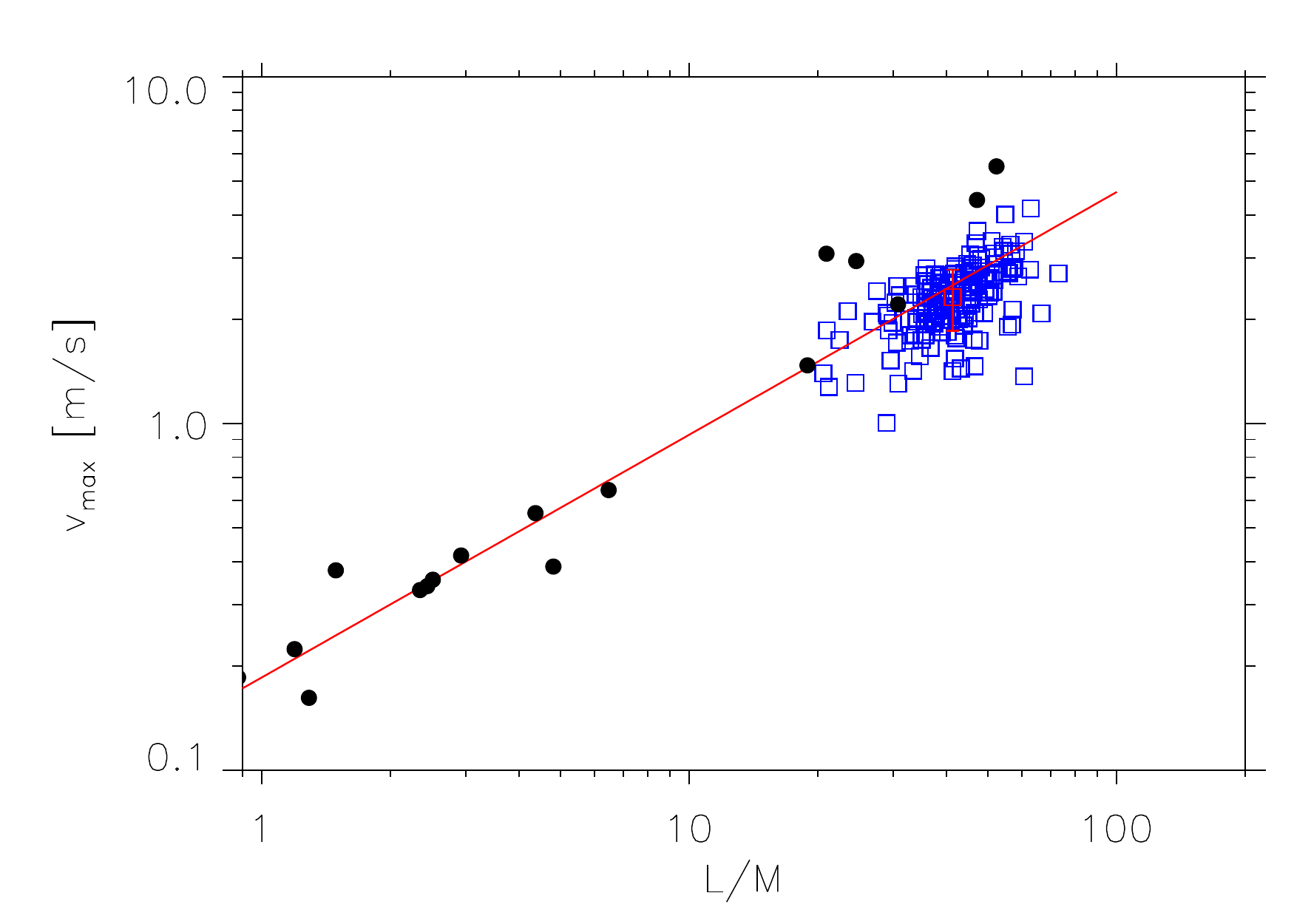}}
         \resizebox{\hsize}{!}{\includegraphics  {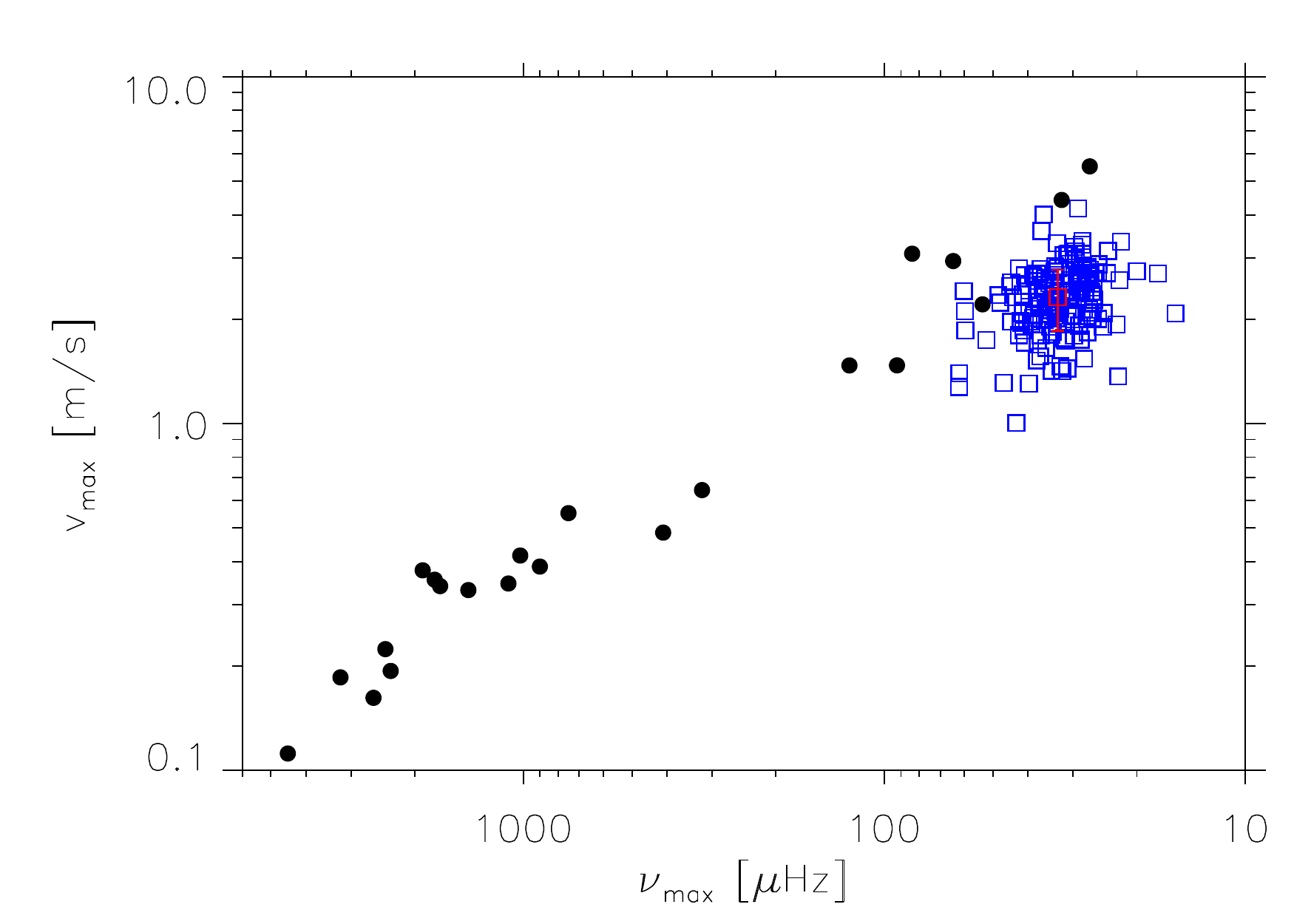}}
       \end{center}    
        \caption{{\bf Top}: Maximum of the mode velocity  $\vmax$ as a function of $L/M$.  The filled circles correspond to the MS stars observed in Doppler velocity from the ground and the red line to the power law of the form $(L/M)^{0.7}$ obtained by \citet{Samadi07a} using 3D models of MS stars. The blue squares correspond to the theoretical $\vmax$ derived according to \eq{eq:vmax} (see Sect.~\ref{comp_velocity}). The red square corresponds to the median value of the theoretical $\vmax$ and the associated vertical bar corresponds to bias introduced  by the 1-$\sigma$ error associated with  the parameters $p$, $s$, ${\cal P}_0$, and ${\cal M}_0$ (Eq.~\ref{scaling_PM}). {\bf Bottom:} Same as top as a function of $\nu_{\rm max}$.}
        \label{scaling_v}
        \end{figure}

Theoretical values of $\vmax$  were compared with the  stars whose  $\vmax$ has been
measured so far in Doppler velocity from the ground.  We considered the
different measurements published in the literature
\citep{Frandsen02,Barban04,Bouchy05,Carrier05,Carrier05b,Mosser05,Arentoft08,Kjeldsen08,Mosser08,Teixeira09,Ando10}. The
values quoted in the literature are generally given in terms of peak
amplitudes. In that case they were converted into
\emph{root-mean-square} (rms) amplitudes. Furthermore, we rescaled  all
amplitudes into \emph{intrinsic} (by opposition to observed)
amplitudes.
Measured values of  $\vmax$ are shown in
Fig.~\ref{scaling_v} (top panel) as a function of $L/M$.  We have an estimate of the ratio $L/M $ for only a few stars  while for almost all of them we have a seismic measure of $\nu_{\rm max}$, which is typically more accurate than the determination of the ratio $L/M$. Therefore, we also show    $\vmax$ in Fig.~\ref{scaling_v} (bottom) as a function of  $\nu_{\rm max}$. 
The theoretical values of $\vmax$ obtained  for our subset of red giants   are found to be close to the measurements obtained for the  red giant stars observed in Doppler velocity  from the ground. 
Note that the considerable dispersion seen in the theoretical values of $\vmax$ comes from the dispersion in the measured value of $\tau_{\rm max}$. 
 Furthermore, we point out that the parameters  $p$, $s$, ${\cal P}_0$, and ${\cal M}_0$, which appear in \eq{scaling_PM}, are mostly determined with quite a large error.  The errors  associated with the  parameters introduce a  bias on the theoretical $\vmax$, which is shown in Fig.~\ref{scaling_v} by a red vertical bar.
 As seen in Fig.~\ref{scaling_v}, the theoretical  $\vmax$ are found, on average, to be about 30\,\% lower than the measurements. 


Using several 3D simulations of the surface of main-sequence stars,
\citet{Samadi07a} have found that $\vmax$ scales as $(L/M)^{sv}$
with $sv=0.7$.  As seen in Fig.~\ref{scaling_v}, this scaling law
reproduces the  MS stars quite well. 
When extrapolated to the red giant domain ($L/M \gtrsim 10 ~ L_\odot/M_\odot$), this scaling
law results for $\vmax$ in values very close to our present theoretical calculations.

The mode masses ${\cal M}_{\rm max}$ were so far evaluated a the reference optical depth $\tau_{\rm~500~nm} = 0.013$ (see Sect.~\ref{velocity}).
 We now discuss the sensitivity ${\cal M}_{\rm max}$ to the  optical depth at which they are  computed.
To evaluate our sensitivity to this choice, we alternatively computed  the theoretical $\vmax$ at the photosphere and at an optical depth ten times lower than our reference level, that is  at $\tau_{\rm~500~nm} = 10^{-3}$. Theoretical $\vmax$ are found to be $\sim $ 30\,\% lower at the photosphere and higher by $\sim 20\,\%$ at the optical depth  $\tau_{\rm~500~nm} = 10^{-3}$.   This result illustrates at which level  $\vmax$ is sensitive to the depth where the acoustic modes are supposed to be measured. This depth is  not well known, however, but we believe that it should be located between the photosphere and  our reference optical depth.

\subsection{Maximum bolometric amplitude  ($(\delta L/L)_{\rm max}$)}
\label{comp_intensity}

\subsubsection{Adiabatic case}
\label{comp_intensity_ad}

\begin{figure}
\begin{center}
 \resizebox{\hsize}{!}{\includegraphics  {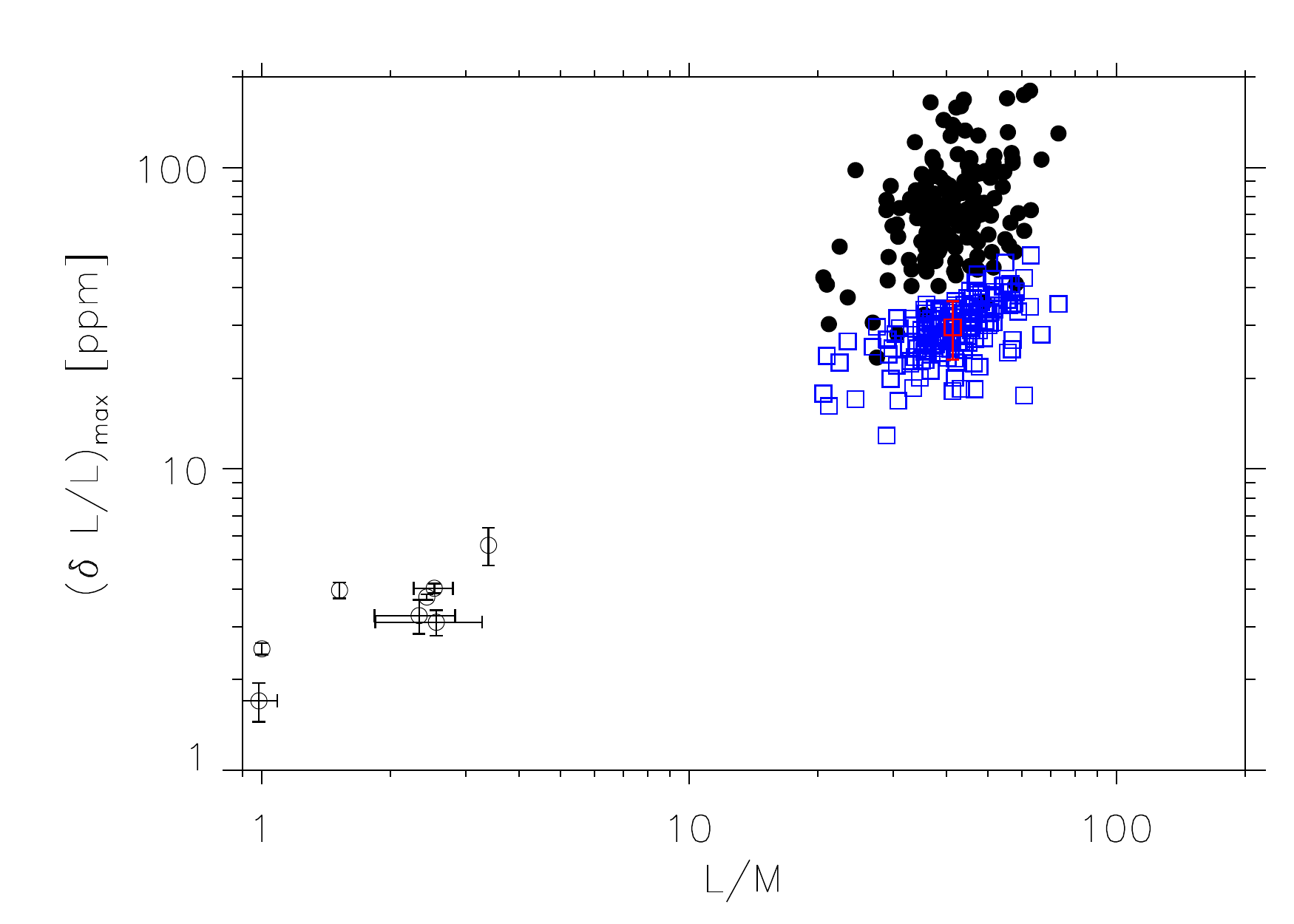}}
 \resizebox{\hsize}{!}{\includegraphics  {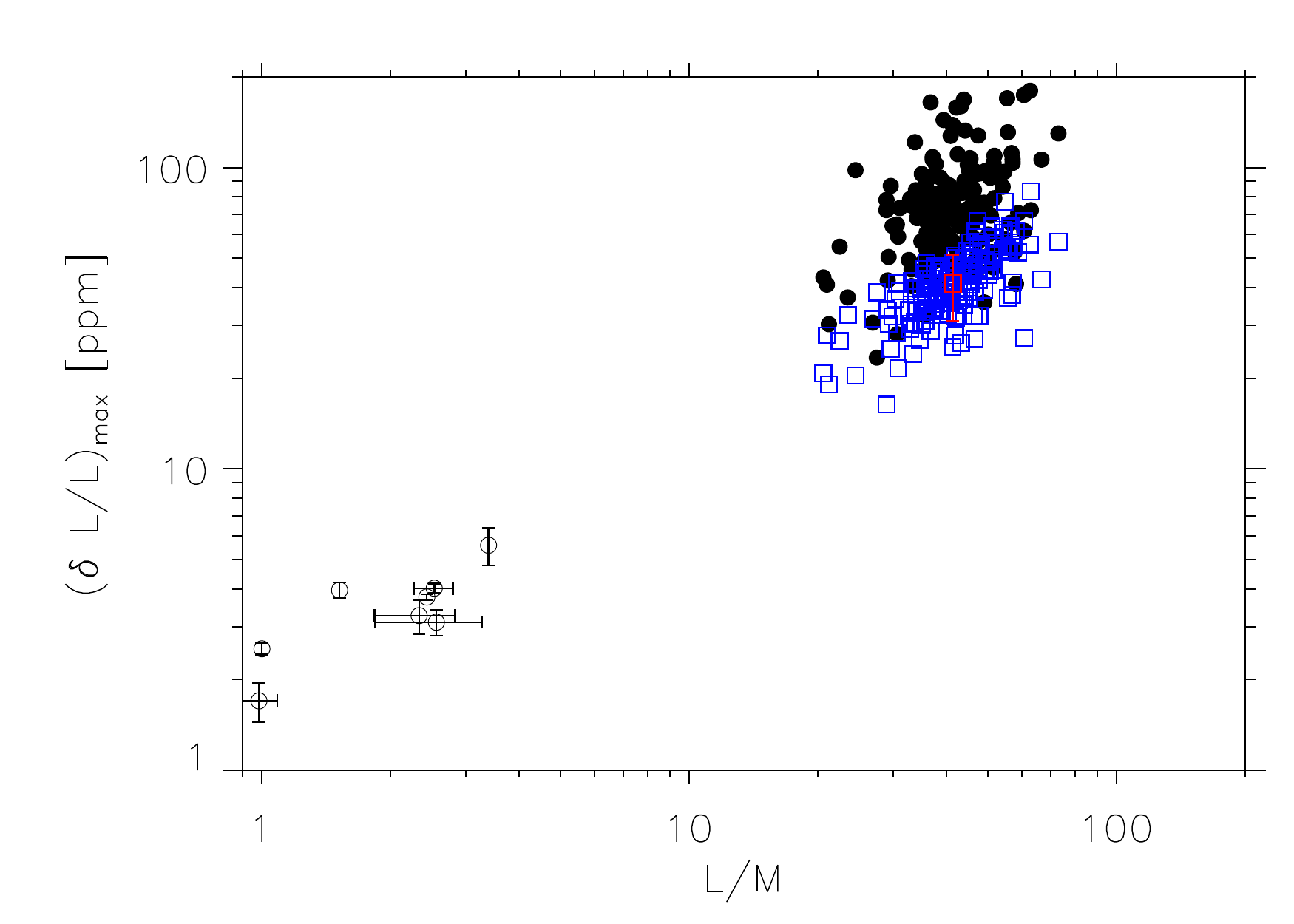}}
\caption{{\bf Top}: Maximum of the mode intensity fluctuation $(\delta L/L)_{\rm max}$ as a function of $L/M$. The filled circles correspond to the seismic measurements performed by \citet{Baudin11} on a large number of CoRoT red giant stars ($\sim 170$ targets). We  only considered the targets for which the mode line width is broader than twice the frequency resolution (see Sect.~\ref{corot_data}). The empty circles correspond to the MS stars observed so far by CoRoT \citep[see][]{Baudin11}, and  the blue squares are  the theoretical $(\delta L/L)_{\rm max}$  computed according to the \citet{Kjeldsen95} adiabatic coefficient (\eq{zeta_K95}, see Sect.~\ref{comp_intensity_ad}).  The red square corresponds to the median value of the theoretical $\dLmax$ and the associated vertical bar corresponds to the bias introduced by the 1-$\sigma$ error associated with the parameters $p_*$, $s$, ${\cal P}_0$, and ${\cal M}_{0,*}$,  $\dLrms^\odot$ and $v_{\rm rms}^\odot$ (see \eq{dLmax}). {\bf Bottom:} Same as top, the  theoretical  $(\delta L/L)_{\rm max}$ are computed here assuming for $\zeta$ the non-adiabatic scaling relation established in Sect.~\ref{nad} (see also Fig.~\ref{intensity-velocity}). The red error bar here also accounts for the 1-$\sigma$ error associated with the parameters $k$ and $\zeta_0$ (see \eq{zeta_nad} and Sect.~\ref{nad}). }
\label{scaling_dL}
\end{center}
\end{figure}

We computed  $(\delta L/L)_{\rm max}$ according to \eq{dLmax}  using  the scaling law given by \eq{eq:vmax} for $v$ and assuming the adiabatic coefficient $\zeta_{\rm K95}$ (\eq{zeta_K95}).
Fig.~\ref{scaling_dL} (top) shows $(\delta L/L)_{\rm max}$  as a function of ratio $(L/M)$, where this ratio is estimated according to \eq{L_M}.
We  also plotted the mode amplitudes  measured for a small sample of CoRoT main-sequence stars \citep[see][and references therein]{Baudin11}.
Theoretical $(\delta L/L)_{\rm max}$ under-estimates the amplitudes measured on the  CoRoT red giant stars by a factor of about 2.5.

\subsubsection{Non-adiabatic case}
\label{comp_intensity_nad}

We computed   $(\delta L/L)_{\rm max}$ according to \eq{dLmax}  assuming  the non-adiabatic scaling law  established in Sect.~\ref{nad} (see \eq{zeta_nad}) for $\zeta$. 
The result is shown in Fig.~\ref{scaling_dL} (bottom). 
Using the non-adiabatic coefficient results in an increase of the bolometric amplitude by a factor $\sim 1.5$ compared to the calculations based on the adiabatic coefficient. This renders the theoretical bolometric amplitude  closer to the observations. 

 We have plotted in Fig.~\ref{diff_dL} the histogram of the relative difference between observed and theoretical $\dLmax$, that is, the histogram of the quantity $\gamma \equiv (A^{obs}-A)/A$, where $A$ is the theoretical amplitude and $A^{obs}$ the observed one. The dispersion seen in the histogram is due both to the errors associated with the data and the fact that we observe a heterogeneous population of stars with different chemical abundance.

The red horizontal  bar shows the  bias introduced by the  1-$\sigma$ errors associated with the determination of the  parameters $p_*$, $s$, ${\cal P}_0$, ${\cal M}_{0,*}$, $k$, and $\zeta_0$ as well the measurement of $\dLrms^\odot$ and $v_{\rm rms}^\odot$ (see \eq{dLrms}).
The median of $\gamma$ is close to 0.8 (the vertical dashed line). This means that  theoretical amplitudes remains, on average, $\sim$\,40\,\% below the CoRoT measurements.

\fig{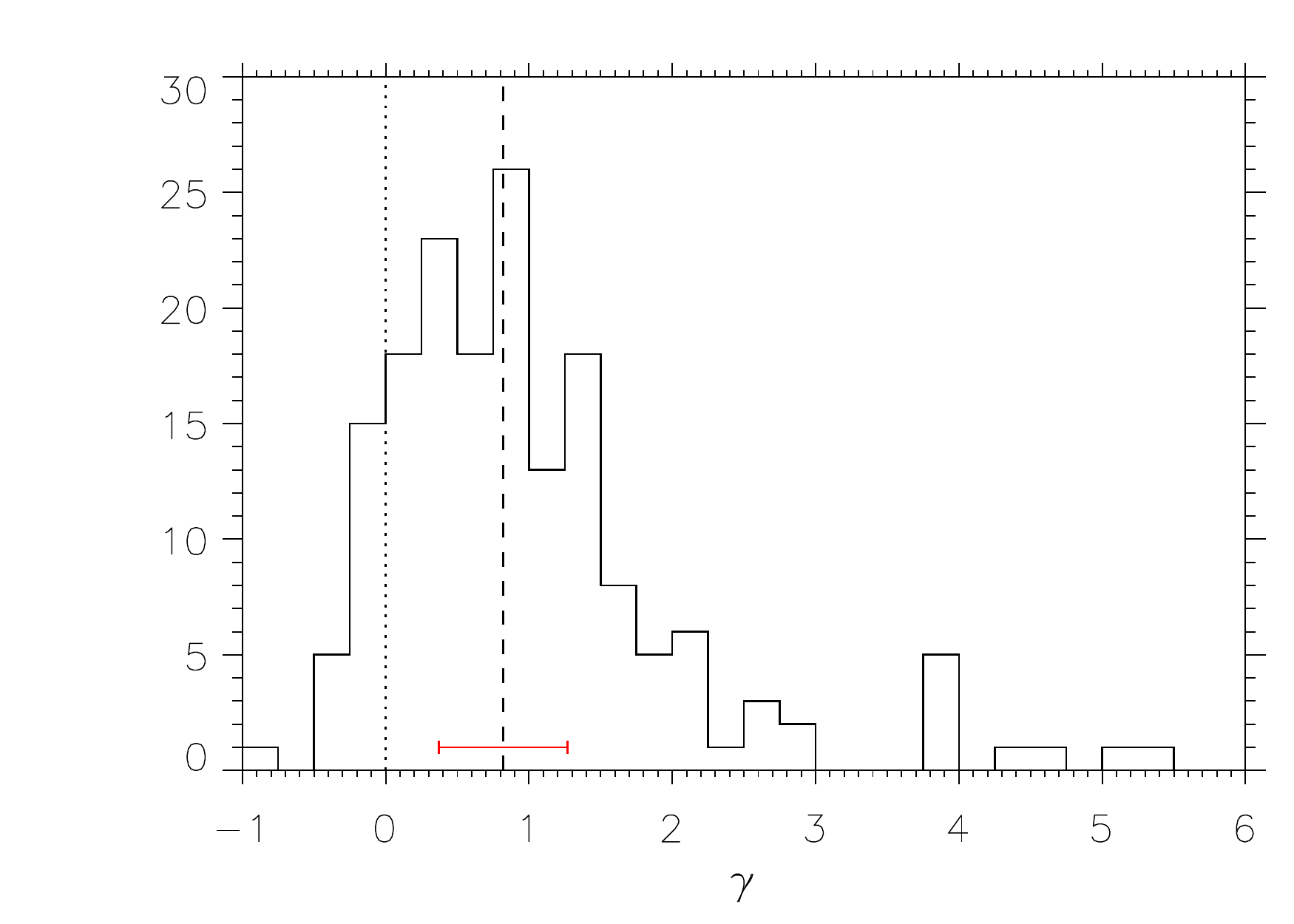}{ Histogram of the relative difference ($\gamma$) between observed and theoretical $\dLmax$ (see text). The vertical dashed line shows the position of the median value. The horizontal error bar  corresponds to the bias introduced by the 1-$\sigma$ error associated with the determination of the parameters $p_*$, $s$, ${\cal P}_0$, ${\cal M}_{0,*}$, $k$, and $\zeta_0$ and the measurement of $\dLrms^\odot$ and $v_{\rm rms}^\odot$.}{diff_dL}

\section{Conclusion}
\label{conclusion}

\subsection{Theoretical scaling relation for the velocity mode amplitude}

We have extended the calculations performed by \citet{Samadi07a} for main-sequence stars to  sub- and red giant stars. 
We found  that the maximum of the  mode excitation rate, $\cal P_{\rm max}$, scales approximately as $(L/M)^s$ with $s = 2.60 \pm 0.08$. 
Accordingly, for  sub- and red giant stars, theoretical $\cal P_{\rm max}$ scales in same way as for the main-sequence stars.

 We also found that the mode mass at the peak frequency, $\cal M_{\rm max}$, which was evaluated at a reference level in the atmosphere, scales as $ \Delta \nu^{-p}$ where $\Delta  \nu \propto (M/R^3)^{1/2}$, with $ p = 2.1 \pm 0.1 $. 
Since $ (M/R^3)$  represents also the mean density, we have that $\cal M_{\rm max}$ scales almost linearly as the inverse of  the star mean density. 
This tight relation still remains to be understood, however. 

From the scaling laws for  $\cal M_{\rm max}$ and $\cal P_{\rm max}$, we finally derived  a scaling law for the maximum of the mode velocity, which has the following form: 
\eqn{
\vmax \propto  \left ( \tau_{\rm max} \right ) ^{1/2}   \,  \left ( {L} \over {M} \right )^{s/2} \, \left ( M \over R^3 \right )^{p/4} \, , 
\label{eq:vmax:2}
}
where $\tau_{\rm max}$ is the mode lifetime at the peak frequency. 

 Using CoRoT data, \citet{Baudin11} have found that $ \tau_{\rm max}$ scales approximately as $\teff^{-m}$ where $m = 16.2 \pm 2 $ for the main-sequence and sub-giant stars. 
Recently,  \citet{Appourchaux12} have found a slope  $m = 15.5 \pm 1.6 $ with {\it Kepler} data, which is hence compatible with that of \citet{Baudin11}. Such a power law is also supported by the theoretical calculations of \citet{Kevin12} performed for main-sequence, sub- and red giant stars. Furthermore, although $\cal M_{\rm max}$ scales better with $ \Delta \nu$, it also  scales   well as $g^{-p^\prime}$ with $p^\prime = 1.66 \pm 0.15$ (note the larger uncertainty for $p^\prime$ compared to $p$).   Accordingly, since $L/M \propto \teff^4/g$,  we can rewrite  the scaling for $\vmax$ (Eq.~\ref{eq:vmax:2}) as a function of the star spectroscopic parameters only:
\eqn{
\vmax \propto \teff  ^{\left (2 s  -m/2 \right ) }   \,   g  ^{\left ( p^\prime/2-s/2 \right )}  \; .
\label{eq:vmax:3}
}

Using a set of CoRoT red giant stars for which the mode lifetimes have been measured \citep{Baudin11}, we derived from the scaling law of Eq.~(\ref{eq:vmax:2}) theoretical values of $\vmax$.  These values were found to be close to  the measurements made from the ground in terms of Doppler velocity for red giant stars. 
However,  the Doppler measurements remain on average under-estimated by a about $30$\,\%. 
We discuss in Sect.~\ref{discussion} possible reasons for this under-estimation.

\subsection{Theoretical scaling relation for the bolometric mode amplitude}

When converted in terms of intensity using the \citet{Kjeldsen95} adiabatic relation, the theoretical amplitudes under-estimate the bolometric mode amplitudes measured by \citet{Baudin11} on a set of CoRoT red giant stars by a factor about 2.5.
Alternatively,  we have considered the  MAD non-adiabatic pulsation code \citep{Grigahcene05} to establish a non-adiabatic relation between  intensity and velocity. We found that this  relation scales as  $(L/M)^k$ with $k = 0.25 \pm 0.05$. We finally established for the mode amplitude in \emph{intensity}  the following scaling law:
\eqn{ \dLmax  \propto  \, \left ( \tau_{\rm max} \right ) ^{1/2}   \,  \left ( {L} \over {M} \right )^{s/2 + k} \, \left ( M \over R^3 \right )^{p_*/4} \, ,
\label{dLmax:2}
}
where $p_* =  2.0 \pm 0.1 $. 
As for the scaling relation for $\vmax$, the one for  $\dLmax $ can be rewritten as  a function of the star spectroscopic parameters only:
\eqn{ 
\dLmax  \propto  \,  \teff  ^{\left (2 s   -m/2  + 4 k  \right )}   \,   g ^{\left ( p_*^\prime/2-s/2 -k\right )}  \;,
\label{dLmax:3}
}
where $p_*^\prime = 1.63 \pm 0.15$.

Using the non-adiabatic scaling law for $\dLmax$ reduces  the difference between theoretical  and  measured amplitudes by a factor $\sim$~1.5. 
Our analysis hence explains qualitatively the recent results obtained for red giant stars using photometric CoRoT and {\it Kepler} observations \citep{Baudin11,Huber11,Stello11,Mosser12}. Indeed, we stress that theoretical relation obtained for mode amplitudes in velocity \emph{cannot} be simply extrapolated into photometry because non-adiabatic effects dominate the relation between mode amplitude in velocity and intensity. 

However, while the non-adiabatic treatment implemented in the MAD code \citep{Grigahcene05}  reduces the discrepancy with the CoRoT measurements, the latter  are still underestimated on average by about 40\,\%. Possible reasons for  this discrepancy are discussed in Sect.~\ref{discussion}. 


\section{Discussion}
\label{discussion}





The mode masses are sensitive to the layer at which they are evaluated, which  must in principle correspond to the height in the atmosphere at which spectrographs dedicated to stellar seismology are the most sensitive \citep[See Sect.~\ref{velocity}, Sect.~\ref{comp_velocity}, and][]{Samadi08}.
 However, the uncertainty associated with the lack of knowledge of this layer introduces an uncertainty on the computed amplitudes that should not exceed $\sim$\,30\,\% (see Sect.~\ref{comp_velocity}). 

The discrepancy with the velocity measurements can also be attributed to the  under-estimation of the mode driving.  It is not clear which part of the excitation model might be incorrect or incomplete. Nevertheless, we believe that a possible bias can arise from the way oscillations are currently treated in the region where the driving is the most efficient  (i.e. the uppermost part of the convective region). Indeed, in this region the oscillation period, the thermal  time-scale and the dynamical time-scale are of the same order, making the coupling between pulsation and convection stronger and energy losses more significant \citep[see e.g.][and references therein]{Kevin11}. 
We have compared non-adiabatic and adiabatic eigenfunctions computed for the global standard 1D model. 
 The  non-adiabatic eigenfunctions obtained with the  MAD pulsation code  differ from the adiabatic ones only in a small fraction of the excitation region. We found a negligible  difference between excitation rates computed with  non-adiabatic eigenfunctions  and those computed with adiabatic eigenfunctions.
However, we point out that the underlying theory is based on a time-dependent version of the mixing-length theory, which is well known to be a crude formulation of convection. Therefore a more realistic and consistent non-adiabatic approach that does not rely on free parameters and that includes constraints from 3D hydrodynamical models is  required.  

Finally, part of the differences with amplitudes  $\dLmax$  measured by CoRoT can be attributed to the intensity-velocity relation. Indeed, if we suppose that the mode masses are correct, then we must multiply the mode excitation rates $\pmax$ by a factor $\sim 1.5^2 = 2.25 $ to match the velocity measurements.
In that case  only a difference of about 20\,\% with the observed $\dLmax$ remains, which must then be  attributed to the intensity-relation.
The intensity-relation strongly depends on the way non-adiabatic  effects are treated, and  as mentioned above, the current non-adiabatic treatment  is based on a crude description of the convection and its inter-action with pulsation.  



\begin{acknowledgements}
The CoRoT space mission, launched on
December 27 2006, has been developed and is operated by CNES,
with the contribution of Austria, Belgium, Brasil, ESA, Germany
and Spain.
\end{acknowledgements}

\bibliographystyle{aa}


\end{document}